\documentclass{article}
\usepackage{graphicx}
\usepackage{amsmath}

\input{tcilatex}

\begin{document}

\title{Propagating Quantum Walks: the origin of interference structures.}
\author{Peter L. Knight\thanks{%
E-mail: p.knight@imperial.ac.uk}, \\
%EndAName
Optics Section, Blackett Laboratory, Imperial College London, \\
London SW7 2AZ, United Kingdom \and Eugenio Rold\'{a}n\thanks{%
E-mail: eugenio.roldan@uv.es}, \\
%EndAName
Departament d'\`{O}ptica, Universitat de Val\`{e}ncia, \\
Dr. Moliner 50, 46100--Burjassot, Spain \and J. E. Sipe\thanks{%
E-mail: sipe@physics.utoronto.ca} \\
%EndAName
Department of Physics, University of Toronto, \\
Toronto M5S 1A7, Canada. }
\maketitle

\begin{abstract}
We analyze the solution of the coined quantum walk on a line. First, we
derive the full solution, for arbitrary unitary transformations, by using a
new approach based on the four "walk fields" which we show determine the
dynamics. The particular way of deriving the solution allows a rigorous
derivation of a long wavelength approximation. This long wavelength
approximation is useful as it provides an approximate analytical expression
that captures the basics of the quantum walk and allows us to gain insight
into the physics of the process.
\end{abstract}

\section{Introduction}

Consider a two--state particle, \textit{e.g.} a two--level atom, moving
along a one--dimensional lattice. The displacement is conditional; that is,
the particle jumps one site at each time to the right or to the left
depending on its internal state. The movement of such a particle is rather
trivial: If the particle is in its \textquotedblright up\textquotedblright\
or \textquotedblright down\textquotedblright\ internal state it moves to the
right or to the left, respectively, and it undergoes a superposition of
right and left displacements if it is in a superposition of its
\textquotedblright up\textquotedblright\ and \textquotedblright
down\textquotedblright\ states. But the movement becomes nontrivial if after
each displacement a suitable unitary transformation, \textit{e.g.} the
Hadamard transformation, leads to a superposition of the internal states of
the particle. Now the probability distribution that the particle would be
found at a given position is quite rich: It is null at odd (even) positions
at odd (even) times, shows two strong peaks near the right and left edges,
develops strong oscillations near these peaks and has a smooth plateau in
the central region. This process is known as the (coined) quantum walk.

In a sense, the quantum walk, which we refer to as QW, is the quantum
generalization of the classical random walk, and as such was first
introduced ten years ago by Aharonov \textit{et al.} \cite{Aharonov}. It was
later rediscovered by Meyer \cite{Meyer} in the context of quantum cellular
automata, and more recently by Watrous \cite{Watrous} in the context of
quantum algorithms. The QW introduced in these papers is called the \textit{%
discrete} QW (be it coined \cite{Aharonov,Meyer} or not \cite{Watrous}), and
must be distinguished from the so--called \textit{continuous} QW, introduced
by Farhi and Gutmann in 1998 \cite{Farhi}, which realizes to some extent a
quantum version of continuous time classical Markov chains. The connection
between these two types of QWs is still unclear \cite{Kempe2}.

The QW, both in its discrete and continuous versions, is a process that has
recently attracted the attention of many workers in the quantum computing
community; see, for example, the recent review by Julia Kempe [5]. Of
interest is the possibility that QWs will give rise to new quantum
algorithms that show clear speedups over classical algorithms. The first
results in this direction have been obtained by Shenvi \textit{et al.} \cite%
{Shenvi}, who have shown that a QW can perform the same tasks as Grover's
search algorithm \cite{Grover}, and by Childs \textit{et al.} \cite{Childs},
who have introduced an algorithm for crossing a special type of graph
exponentially faster than can be done classically.

In the present communication we give an alternate description of the
discrete QW that we feel gives physical insight into the nature of that
process. This paper is an extention of our previous work \cite{Knight}, in
which we showed that the QW on the line is not a characteristically quantum
process in the specific sense that it can be understood simply as a wave
phenomenon, thus opening the possibility of entirely classical
implementations \cite{Knight2}. Indeed, wave implementations \cite%
{Hillery,Jeong} of the QW have already been discussed. In other words, the
characteristic quantum feature of entanglement does not seem to play any
central role in the QW, at least in the unidimensional case. The situation
could be different for higher dimensional QWs; see the discussion in Knight 
\textit{et al. }\cite{Knight2} for some insight into the role of
entanglement in higher dimensional QWs.

The rest of the article is organized as follows. In Section 2 we give the
evolution equations for the standard discrete coined QW, and reformulate
them in a way that directly exhibits an important property: The two coin
states evolve independently of each other, the coupling between them
entering only through the first iteration of the walk. In Section 3 we
derive the formal solution of the QW in a manner different to those already
carried out \cite{Nayak,Carteret}. This alternative derivation allows a
rigorous derivation of the long wavelength approximation that corresponds to
the continuous limit of the QW we pointed out earlier \cite{Knight}. In
Section 4 we give the solution of this long wavelength limit and discuss
some of its properties. Finally in Section 5 we give our conclusions.

\section{Equations of evolution}

Let $\mathcal{H}_{P}$ be the Hilbert space of the particle positions and $%
\left\{ \left\vert m\right\rangle ,m\in Z\right\} $ a basis of $\mathcal{H}%
_{P}$; and let $\mathcal{H}_{C}$ be the Hilbert space describing the two
states of the qubit associated with the internal states of the particle
(usually referred to as the coin), and $\left\{ \left\vert R\right\rangle
,\left\vert L\right\rangle \right\} $ a basis of $\mathcal{H}_{C}$. The
state of the total system belongs to the space $\mathcal{H}=\mathcal{H}%
_{C}\otimes \mathcal{H}_{P}$. The dynamics of the system is governed by two
operations, the conditional diplacement 
\begin{eqnarray}
\hat{D}\left\vert m,R\right\rangle  &=&\left\vert m+1,R\right\rangle , \\
\hat{D}\left\vert m,L\right\rangle  &=&\left\vert m-1,L\right\rangle ,
\end{eqnarray}%
and the transformation acting on the internal states of the coin 
\begin{eqnarray}
\hat{H}\left\vert m,R\right\rangle  &=&\sqrt{\rho }\left\vert
m,R\right\rangle +\sqrt{1-\rho }\left\vert m,L\right\rangle , \\
\hat{H}\left\vert m,L\right\rangle  &=&\sqrt{1-\rho }\left\vert
m,R\right\rangle -\sqrt{\rho }\left\vert m,L\right\rangle ,
\end{eqnarray}%
$0\leq \rho \leq 1$, which is the most general unitary transformation that
one needs to consider \cite{Bach} (notice that $\hat{H}$ is the Hadamard
transformation when $\rho =1/2$). The QW is implemented by the repeated
action of the operator $\hat{H}\hat{D}$, \textit{i.e.} if $\left\vert \psi
\right\rangle _{n}$ denotes the state of the system after $n$ iterations
then 
\begin{equation}
\left\vert \psi \right\rangle _{n}=\left( \hat{H}\hat{D}\right)
^{n}\left\vert \psi \right\rangle _{0},
\end{equation}%
which can be written as 
\begin{equation}
\left\vert \psi \right\rangle _{n}=\sum_{m=-n}^{+n}\left[ R_{m,n}\left\vert
m,R\right\rangle +L_{m,n}\left\vert m,L\right\rangle \right] ,
\end{equation}%
with the equations of evolution of the probability amplitudes given by 
\begin{eqnarray}
R_{m,n} &=&\sqrt{\rho }R_{m-1,n-1}+\sqrt{1-\rho }L_{m+1,n-1},  \label{amp1}
\\
L_{m,n} &=&\sqrt{1-\rho }R_{m-1,n-1}-\sqrt{\rho }L_{m+1,n-1}.  \label{amp2}
\end{eqnarray}%
Finally, the probability of finding the particle at position $m$ at
iteration (time) $n$, is given by 
\begin{eqnarray}
P_{m}\left( n\right)  &=&P_{m}^{R}\left( n\right) +P_{m}^{L}\left( n\right) 
\label{P} \\
P_{m}^{R}\left( n\right)  &=&\left\vert R_{m,n}\right\vert
^{2},\;\;\;\;P_{m}^{L}\left( n\right) =\left\vert L_{m,n}\right\vert ^{2}. 
\notag
\end{eqnarray}

Eqs.(\ref{amp1}) and (\ref{amp2}) are the discrete QW equations as can be
found \textit{e.g.} in the work of Carteret \cite{Carteret} for the specific
case of the fair coin $\rho =1/2$. Different equations of evolution, but
still leading to the same final probability, can be found by defining the
walk as the repeated action of the operator $\hat{D}\hat{H}$, instead of $%
\hat{H}\hat{D}$, as is done by Nayak and Vishwanath  \cite{Nayak}. In the
Appendix we identify the transformations that connect these different
conventions for the probability amplitudes.

There is a crucial feature in Eqs.(\ref{amp1}) and (\ref{amp2}) that is at
the heart of our treatment of the QW. Substituting Eq.(\ref{amp2}) into Eq.(%
\ref{amp1}) and rearranging the terms one easily gets 
\begin{equation}
R_{m,n}=\sqrt{\rho }R_{m-1,n-1}+R_{m,n-2}-\sqrt{\rho }\left( \sqrt{\rho }%
R_{m,n-2}+\sqrt{1-\rho }L_{m+2,n-2}\right) .
\end{equation}%
Noticing that the last term in the right--hand side is nothing but $%
R_{m+1,n-1}$, and making the change $n\rightarrow n+1$, we can write 
\begin{equation}
R_{m,n+1}=R_{m,n-1}+\sqrt{\rho }R_{m-1,n}-\sqrt{\rho }R_{m+1,n}.
\end{equation}%
A similar equation for $L_{m,n}$ can be derived. That is, Eqs.(\ref{amp1})
and (\ref{amp2}) are equivalent to 
\begin{equation}
a_{m,n+1}-a_{m,n-1}=\sqrt{\rho }\left[ a_{m-1,n}-a_{m+1,n}\right]
,\;\;\;a=R,L  \label{amp3}
\end{equation}%
which is a remarkable equation, as it shows that the evolutions of the two
coin states are independent. That is, $R_{m,n}$ does not depend on $L_{m,n}$
at previous times, and \textit{vice versa. }The only coupling between $%
R_{m,n}$ and $L_{m,n}$ appears at the first interation. Notice that in Eq. (%
\ref{amp3}) one needs to specify two initial conditions (both $a_{m,0}$ and $%
a_{m,1}$) and one needs Eqs.(\ref{amp1}) and (\ref{amp2}) for calculating $%
a_{m,1}$ from $a_{m,0}$. But apart from this initial coupling, the evolution
is completely independent for the two coin states. In other words: The QW is
composed of two independent QWs, one for each coin state, coupled only at
initialization.

Eq.(\ref{amp3}) is interesting for another reason already discussed in \cite%
{Knight} in the special case of the Hadamard transformation ($\rho =1/2$).
Simple inspection leads to the idea that Eq.(\ref{amp3}) can be understood
as the discretization of a differential equation involving all the odd
derivatives with respect to space and time of a continuous field $a\left(
x,t\right) $%
\begin{equation}
\sum_{k=0}^{\infty }\frac{\Delta t^{2k+1}}{\left( 2k+1\right) !}\frac{%
\partial ^{2k+1}}{\partial t^{2k+1}}a\left( x,t\right) =-\frac{1}{\sqrt{2}}%
\sum_{k=0}^{\infty }\frac{\Delta x^{2k+1}}{\left( 2k+1\right) !}\frac{%
\partial ^{2k+1}}{\partial x^{2k+1}}a\left( x,t\right) .  \label{aprox}
\end{equation}%
This is too na\"{\i}ve an approximation, as the above equation does not
respect the symetries of the solutions of Eq.(\ref{amp3}): Eq.(\ref{aprox})
describes the propagation of waves in the positive position \textrm{x}
direction, whilst the probability distribution of the quantum walk clearly
shows propagation in both the positive and negative directions. But it is
clear that Eq.(\ref{amp3}) suggests the existence of some kind of continuous
description of the discrete coined QW. Earlier \cite{Knight} we obtained
such a continuous description by introducing, without justification, new
amplitudes $A_{m,n}^{\pm }$ through the definition

\begin{equation}
a_{m,n}=A_{m,n}^{+}+\left( -1\right) ^{n}A_{m,n}^{-},  \label{ansatz}
\end{equation}%
and then we derived continuous differential equations for the continuous
fields $A^{\pm }\left( x,t\right) $. These equations are similar to Eq.(\ref%
{aprox}) but with a velocity that differs in its sign for the fields $%
A^{+}\left( x,t\right) $ and $A^{-}\left( x,t\right) $. From these equations
we obtained an approximate evolution equation by neglecting higher order
derivatives 
\begin{equation}
\frac{\partial }{\partial t}A^{\pm }\left( \xi ,\tau \right) =\mp \frac{1}{%
\sqrt{2}}\left[ \frac{\partial }{\partial \xi }+\frac{1}{12}\frac{\partial
^{3}}{\partial \xi ^{3}}\right] A^{\pm }\left( \xi ,\tau \right) ,
\label{aprox2}
\end{equation}%
with $\xi =x/\Delta x$ and $\tau =t/\Delta t$. Eq.(\ref{aprox}) can be
solved explicitly in terms of Airy functions, and shows that to this degree
of approximation the quantum walk fields $A_{m,n}^{\pm }$ satisfy the same
equation as the evolution of an optical pulse in a medium with third order
dispersion.

The weak points of this approach \cite{Knight} are that the fields $A^{\pm }$
are introduced without justification, that only the special case of the
Hadamard transformation is considered, and that the continuum limit is not
clearly derived. The goal of this paper is to present a rigorous derivation
of Eq.(\ref{aprox}) for arbitrary values of $\rho $. For that purpose, we
first turn to a new derivation of the full solution of the quantum walk on
the line.

\section{The exact solution: an approach based on four fields}

As at a given time identified by the index $n$ the amplitudes $a_{m,n}$
depend on the discrete variable $m$. So it is usual \cite{Nayak,Carteret} to
define a continuous function of $\kappa $, periodic over a range of $2\pi $,
such that the $a_{m,n}$ are the Fourier components of that function $%
f_{n}(\kappa )$,%
\begin{equation}
f_{n}(\kappa )\equiv \sum_{m=-\infty }^{\infty }a_{m,n}e^{im\kappa }.
\label{fexpan}
\end{equation}%
It is then possible to formulate the problem in terms of the continuous
function $f_{n}(\kappa )$ of $\kappa $ rather than the discrete function $%
a_{m,n}$ of $m$. A similar strategy can be implemented with respect to the
dependence on the time step $n$. We essentially follow this approach below,
but our perspective is slightly different. Because we want to be able to
understand Eq. (\ref{amp3}) as the discretization of certain linear partial
differential equations, we introduce Fourier expansions such as (\ref{fexpan}%
) in a more physical way.

\subsection{Fourier amplitudes}

From the amplitudes $a_{m,n}$ defined at discrete points in space, labeled
by $m$, and time, labeled by $n$, we can formally introduce a field $a(x,t)$
at arbitrary points in space and time defined by 
\begin{equation}
a\left( x,t\right) =\sum_{m,n=-\infty }^{\infty }a_{m,n}\delta \left(
x-mX\right) \delta \left( t-nT\right) ,  \label{aux0}
\end{equation}%
which produces appropriately weighted delta functions at the positions in
space and time where\thinspace $a_{m,n}$ is defined. Here $X$ is the
distance between the sites in the one--dimensional lattice, and $T$ is the
time between steps in the walk. We can formally introduce the Fourier
transform of this function 
\begin{equation}
a(x,t)=\frac{1}{(2\pi )^{2}}\int_{-\infty }^{+\infty }d\tilde{k}\,d\tilde{%
\omega}\,a(\tilde{k},\tilde{\omega})e^{i\tilde{k}x}e^{-i\tilde{\omega}t},
\end{equation}%
where from Eq.(\ref{aux0}) we find that 
\begin{equation}
a(\tilde{k},\tilde{\omega})\equiv \int_{-\infty }^{+\infty }a(x,t)e^{-i%
\tilde{k}x}e^{i\tilde{\omega}t}=\sum_{m,n=-\infty }^{+\infty }a_{m,n}e^{-i%
\tilde{k}mX}e^{i\tilde{\omega}nT}.  \label{aux1}
\end{equation}%
Next, note that we can write each $\tilde{k}$ and $\tilde{\omega}$ uniquely
as

\begin{equation}
\tilde{k}=k+\frac{2\pi p}{X},\;\;\tilde{\omega}=\omega +\frac{2\pi q}{T},
\end{equation}
with $p$ and $q$ integers, and $k$ and $\omega $ lying within the ranges 
\begin{equation}
-\frac{\pi }{X}<k\leq \frac{\pi }{X},\;\;-\frac{\pi }{T}<\omega \leq \frac{%
\pi }{T}.  \label{range}
\end{equation}
Then using the identity 
\begin{equation}
\sum_{p=-\infty }^{+\infty }\delta (y-2\pi p)=\frac{1}{2\pi }\sum_{p=-\infty
}^{+\infty }e^{ipy},  \label{didentity}
\end{equation}
which follows immediately from identifying the Fourier sum representing the
periodic function of $y$ on the left-hand side, we easily find from Eq.(\ref%
{aux1}) that $a(\tilde{k},\tilde{\omega})=a\left( k,\omega \right) $. Then,
making use of Eqs.(\ref{aux0})--(\ref{aux1}) and (\ref{didentity}), we can
write 
\begin{equation}
a\left( x,t\right) =\bar{a}\left( x,t\right) \sum_{m,n=-\infty }^{\infty
}\delta \left( x-mX\right) \delta \left( t-nT\right)  \label{a(x,t)}
\end{equation}
where 
\begin{eqnarray}
\bar{a}\left( x,t\right) &=&\frac{1}{\left( 2\pi \right) ^{2}}XT\int_{-\pi
/X}^{+\pi /X}dk\int_{-\pi /T}^{+\pi /T}d\omega \,a\left( k,\omega \right)
e^{ikx}e^{-i\omega t}  \label{abardef} \\
a\left( k,\omega \right) &=&\sum_{m,n=-\infty }^{\infty
}a_{m,n}e^{-ikmX}e^{i\omega nT},  \label{a(k,omega)}
\end{eqnarray}
\textit{cf. }Eq.(\ref{aux1}).

\subsection{Dispersion relation}

Returning now to the discrete QW Eqs.(\ref{amp3}), we multiply each side by $%
\exp \left( i\omega nT\right) $ $\exp \left( -ikmX\right) $ and sum over $m$
and $n.$ This yields the equation 
\begin{equation}
\sin \omega T=\sqrt{\rho }\sin kX,  \label{disp}
\end{equation}%
which identifies the dispersion relations $\omega =\omega \left( k\right) $.
Because of the range of $k$ and $\omega $ appearing in Eq.(\ref{abardef}),
we need only consider such solutions over the ranges (\ref{range}). There
are then two such solutions. One solution, which we label $\omega _{0}(k)$,
is that for which $\omega \rightarrow 0$ as $k\rightarrow 0$. The second,
which we label $\omega _{1}(k)$, is given by 
\begin{equation}
\omega _{1}\left( k\right) =-\omega _{0}\left( k\right) \pm \frac{\pi }{T},
\label{disp2}
\end{equation}%
for $k$ respectively $>0$ and $<0$. These relations are represented in Fig.1.

\subsection{Separation into two fields}

Solutions of Eqs.(\ref{amp3}) must respect the dispersion relations, and
thus for Eq.(\ref{a(k,omega)}) to represent an actual solution we must have 
\begin{equation}
a\left( k,\omega \right) =\frac{\left( 2\pi \right) ^{2}}{XT}\left[
a^{+}\left( k\right) \delta \left( \omega -\omega _{0}\left( k\right)
\right) +a^{-}\left( k\right) \delta \left( \omega -\omega _{1}\left(
k\right) \right) \right] ,
\end{equation}%
for some amplitudes $a^{+}(k)$ and $a^{-}(k)$. Thus from Eq.(\ref{abardef})
we find 
\begin{equation}
\bar{a}\left( x,t\right) =\int_{-\pi /X}^{+\pi /X}dk\,a^{+}\left( k\right)
e^{ikx}e^{-i\omega _{0}\left( k\right) t}+\int_{-\pi /X}^{+\pi
/X}dk\,a^{-}\left( k\right) e^{ikx}e^{-i\omega _{1}\left( k\right) t}.
\label{abar}
\end{equation}%
Using this expression for $\bar{a}(x,t)$ in Eq.(\ref{a(x,t)}) for $a(x,t)$,
employing Eq.(\ref{disp2}) to write $\omega _{1}(k)$ in terms of $\omega
_{0}(k)$, and recalling that $\exp \left( \pm i\pi n\right) =\left(
-1\right) ^{n}$ we can write 
\begin{eqnarray}
a(x,t) &=&a^{+}(x,t)\sum_{m,n=-\infty }^{+\infty }\delta (x-mX)\delta (t-nT)
\notag \\
&&+a^{-}(x,t)\sum_{m,n=-\infty }^{+\infty }(-1)^{n}\delta (x-mX)\delta
(t-nT),  \label{axtsol}
\end{eqnarray}%
where 
\begin{equation}
a^{\pm }(x,t)=\int_{-\pi /X}^{+\pi /X}dk\,a^{\pm }\left( k\right)
e^{ikx}e^{\mp i\omega _{0}\left( k\right) t}.  \label{axtft}
\end{equation}%
Obviously, we can also write Eq.(\ref{abar}) in the form 
\begin{equation}
\bar{a}(x,t)=a^{+}(x,t)+\left( -1\right) ^{n}a^{-}(x,t),  \label{abar2}
\end{equation}%
which will be used in the next section.

Clearly the fields $a^{+}(x,t)$ and $a^{-}(x,t)$ represent disturbances
propagating (at least for small $k$) to the right and the left respectively,
both with a dispersion relation $\omega _{0}(k)$. The effect of the second
dispersion relation $\omega _{1}(k)$ is to introduce a kind of ``temporally
antiferromagnetic'' behavior, indicated by the $(-1)^{n}$, associated with
the leftward going wave. The solution seems the simplest when formulated in
terms of these two fields $a^{+}(x,t)$ and $a^{-}(x,t)$. Since we have
dynamical equations, Eqs.(\ref{amp3}), for both the $R_{m,n}$ and the $%
L_{m,n}$, we will have four fields in all as $a$ ranges over $R$ and $L$: $%
R^{+}(x,t)$, $R^{-}(x,t)$, $L^{+}(x,t)$, and $L^{-}(x,t)$.

\subsection{Initial conditions}

Eqs.(\ref{amp3}) requires two sets of initial conditions: the value of $%
R_{m,n}$ and $L_{m,n}$ at both $n=0$ and $n=1$. We can now see how these are
built into the fields $a^{\pm }(x,t)$. We begin by noting that from Eqs.(\ref%
{aux0}) and (\ref{axtsol}) we can immediately identify 
\begin{eqnarray}
a_{m,0} &=&a^{+}\left( mX,0\right) +a^{-}\left( mX,0\right) ,  \label{am0} \\
a_{m,1} &=&a^{+}\left( mX,T\right) -a^{-}\left( mX,T\right) .  \label{am1}
\end{eqnarray}%
So to construct the $a^{\pm }(k)$ appropriate for our initial conditions we
must invert Eqs.(\ref{axtft}) to find the $a^{\pm }(k)$ in terms of $a^{\pm
}(x,t)$. This is easily done by focusing on points $x=mX$; writing Eqs.(\ref%
{axtft}) at such points and performing the sums indicated below, we find 
\begin{equation}
a^{\pm }\left( k\right) =\frac{1}{2\pi }Xe^{\pm i\omega _{0}\left( k\right)
t}\sum_{m=-\infty }^{\infty }a^{\pm }\left( mX,t\right) e^{-ikmX}.
\label{aux2}
\end{equation}%
As the left hand side is independent of time, so must be the right hand
side. Writing down these equations for $t=0$ and $t=T$ we can then use Eqs.(%
\ref{am0}) and (\ref{am1}) to solve for the $a^{\pm }(k)$ in terms of the $%
a_{m,0}$ and the $a_{m,1}$. The result is 
\begin{eqnarray}
a^{\pm }\left( k\right) &=&\frac{1}{4\pi }X\sum_{m=-\infty }^{\infty }\frac{%
e^{\pm i\omega _{0}(k)T}a_{m,0}\pm a_{m,1}}{\cos \left[ T\omega _{0}\left(
k\right) \right] }e^{-ikmX}  \notag \\
&=&\frac{1}{4\pi }X\sum_{m=-\infty }^{\infty }\frac{e^{\pm i\omega
_{0}(k)T}a_{m,0}\pm a_{m,1}}{\sqrt{1-\rho \sin ^{2}kX}}e^{-ikmX},
\label{a(k)}
\end{eqnarray}%
where in the second expression we have used the dispersion relation (\ref%
{disp}).

\subsection{Full solution}

The solution of the walk problem subject to arbitary initial conditions can
then be written in the form of Eq.(\ref{axtsol}) or, identifying the
discrete points of interest in space and time, in the form 
\begin{equation}
a_{m,n}=a^{+}(mX,nT)+(-1)^{n}a^{-}(mX,nT),  \label{finres}
\end{equation}%
where $a^{\pm }(x,t)$ are given by Eqs.(\ref{axtft}) with the Fourier
components given by Eq.(\ref{a(k)}). Inserting those Fourier components we
find that we can write the full solution for $a^{\pm }(x,t)$ in terms of a
single Green function $g(x;t)$, 
\begin{equation}
a^{\pm }(x,t)=\sum_{m=-\infty }^{+\infty }g(\pm (x-mX);t-T)a_{m,0}\pm
\sum_{m=-\infty }^{+\infty }g(\pm (x-mX);t)a_{m,1},
\end{equation}%
where 
\begin{equation}
g(x;t)=\frac{X}{4\pi }\int_{-\pi /X}^{\pi /X}dk\frac{e^{ikx}e^{-i\omega
_{0}(k)t}}{\sqrt{1-\rho \sin ^{2}kX}},
\end{equation}%
and we have used the fact that $g(x;-t)=g(-x;t)$. Using the original walk
equations (\ref{amp1}) and (\ref{amp2}) we can then find the $a_{m,1}$ in
terms of the $a_{m,0}$, for $a=R,L$. Then our final solution is given by Eq.(%
\ref{finres}) with 
\begin{eqnarray}
L^{\pm }(x,t) &=&\sum_{m}g(\pm (x-mX);t-T)\,L_{m,0}\mp \sqrt{\rho }%
\sum_{m}g(\pm (x-mX);t)\,L_{m+1,0}  \notag \\
&&\pm \sqrt{1-\rho }\sum_{m}g(\pm (x-mX);t)\,R_{m-1,0}\ \ ,
\end{eqnarray}%
and 
\begin{eqnarray}
R^{\pm }(x,t) &=&\sum_{m}g(\pm (x-mX);t-T)\,R_{m,0}\pm \sqrt{\rho }%
\sum_{m}g(\pm (x-mX);t)\,R_{m-1,0}  \notag \\
&&\pm \sqrt{1-\rho }\sum_{m}g(\pm (x-mX);t)\,L_{m+1,0}\ .
\end{eqnarray}

Earlier, Nayak and Vishwanath \cite{Nayak} formally solved the discrete QW
and analyzed the asymptotic behaviour of the solution. In the Appendix we
show that our formal solution and theirs are the same. More recently,
Carteret \textit{et al.} \cite{Carteret} have reviewed previous results
concerning the asymptotics of the QW and presented new asymptotics. Of
course, all those asymptotic results fully apply to our solution; we do not
pursue this further here.

\section{An approximate approach: the long wavelength limit}

The solution in the form we have constructed leads naturally to a passage to
the continuum limit, which we identify by focusing on the continuous
functions of space and time $\bar{a}(x,t)$, and constructing simplifed
equations for their evolution in the long wavelength limit.

\subsection{Derivation of the equation of evolution in the long wavelength
limit}

Our starting points are Eqs.(\ref{axtft}) and (\ref{abar2}), which we write
again here for clarity:%
\begin{eqnarray}
\bar{a}(x,t) &=&a^{+}(x,t)+\left( -1\right) ^{n}a^{-}(x,t), \\
a^{\pm }(x,t) &=&\int_{-\pi /X}^{+\pi /X}dk\,a^{\pm }\left( k\right)
e^{ikx}e^{\mp i\omega _{0}\left( k\right) t},
\end{eqnarray}%
with $a^{\pm }\left( k\right) $ given by Eqs.(\ref{a(k)}). We consider the
long wavelength limit by neglecting high frequency (and small wavelength)
effects. For that purpose we introduce low--frequency fields 
\begin{equation}
\hat{A}^{\pm }\left( x,t\right) =\int_{-\pi /L}^{+\pi /L}dk\,\mathcal{G}%
\left( kX\right) a_{\pm }\left( k\right) e^{ikx}e^{\mp i\omega _{0}\left(
k\right) t}.  \label{Axt}
\end{equation}%
which in the long wavelength limit take the place of the fields $a^{\pm
}(x,t)$. Here $\mathcal{G}(kX)$ is a function that vanishes at high spatial
frequencies $k$; that is, it plays the role of a low-pass filter. The
cut-off that it provides is assumed to be such that we can make a series
expansion in $\omega _{0}\left( k\right) $. Taking the dispersion relation
Eq.(\ref{disp}) and expanding as a power series one gets 
\begin{equation}
\omega T-\frac{1}{6}\omega ^{3}T^{3}+\ldots =\sqrt{\rho }Xk-\frac{1}{6}\sqrt{%
\rho }X^{3}k^{3}+\ldots   \label{series}
\end{equation}%
Using the lowest order, $\omega T=\sqrt{\rho }Xk$, we approximate $\left(
\omega T\right) ^{3}\approx \left( \sqrt{\rho }Xk\right) ^{3}$ and write $%
\omega \approx \hat{\omega}\left( k\right) $ with 
\begin{equation}
\hat{\omega}\left( k\right) =\sqrt{\rho }\frac{X}{T}k-\frac{1}{6}\sqrt{\rho }%
\left( 1-\rho \right) \frac{X^{3}}{T}k^{3}.
\end{equation}%
Now, we take this aproximate dispersion relation in Eq.(\ref{Axt}) and write 
\begin{equation}
\hat{A}^{\pm }\left( x,t\right) \approx \int_{-\infty }^{+\infty }dk\,a_{\pm
}\left( k\right) \mathcal{G}\left( kX\right) e^{ikx}e^{\mp i\hat{\omega}%
\left( k\right) t},  \label{Axt2}
\end{equation}%
where $\omega _{0}\left( k\right) $ has been approximated by $\hat{\omega}%
\left( k\right) $ and the integration limits have been extended to infinity.
Finally from Eqs.(\ref{Axt2}) it is easy to derive the equations of
evolution of $\hat{A}^{\pm }\left( x,t\right) $. The equations can be
written in the form 
\begin{equation}
\frac{\partial }{\partial t}A^{\pm }\left( \xi ,\tau \right) =\mp \sqrt{\rho 
}\left[ \frac{\partial }{\partial \xi }+\frac{1-\rho }{6}\frac{\partial ^{3}%
}{\partial \xi ^{3}}\right] A^{\pm }\left( \xi ,\tau \right) 
\label{Diff.eq.}
\end{equation}%
where we have introduced the new functions $A^{\pm }\left( \xi ,\tau \right)
=\hat{A}^{\pm }(X\xi ,T\tau )$, and the normalized time and space variables $%
\tau =t/T$ and $\xi =x/X$. Notice that the form of this equation is
independent of the particular form chosen for the cut-off function $\mathcal{%
G}\left( kX\right) $.

This is the differential equation derived earlier (Eq.(15) of \cite{Knight})
for the special case $\rho =1/2$. It can be identified with the equation the
envelope function of a light pulse would satisfy in a description of
propagation through a material medium with no group velocity dispersion at
the lowest order -- such lowest order dispersion would appear through a term
involving a second derivative with respect to $\xi $ -- but with higher
order dispersion described by the third derivative term. The equation is
well known in fiber optics, where it is used to describe pulse propagation
near the zero-dispersion wavelength \cite{Agrawal}. Notice that the
effective \textquotedblleft dispersion coefficient\textquotedblright\ in Eq.
(\ref{Diff.eq.}) is proportional to $(1-\rho )$ and thus, interestingly, can
be varied by modifying the unitary transformation applied in the QW after
each displacement.

\subsection{Solution of the differential equation}

Let us proceed now to analyze the solution of Eq.(\ref{Diff.eq.}). By
Fourier transforming Eq.(\ref{Diff.eq.}) one easily gets 
\begin{equation}
A^{\pm }\left( \xi ,\tau \right) =\int_{-\infty }^{+\infty }dq\,\bar{A}^{\pm
}\left( q,0\right) e^{iq\xi }e^{\mp i\kappa \sqrt{\rho }\left( 1+\frac{%
1-\rho }{6}q^{2}\right) \tau },  \label{aprel}
\end{equation}%
with 
\begin{equation}
\bar{A}^{\pm }\left( q,0\right) =\frac{1}{2\pi }\int_{-\infty }^{+\infty
}d\xi \,A^{\pm }\left( \xi ,0\right) e^{-iq\xi }.  \label{Aq0}
\end{equation}

The initial conditions $A^{\pm }\left( \xi ,0\right) $ can be calculated
from Eqs.(\ref{a(k)}) and (\ref{Axt2}). First we rewrite Eqs.(\ref{a(k)}) in
the long--wavelength limit 
\begin{equation}
a^{\pm }\left( k\right) =\frac{1}{4\pi }X\sum_{m=-\infty }^{\infty }\left(
a_{m,0}\pm a_{m,1}\right) e^{-imkX}
\end{equation}%
where we have approximated $\exp \left[ T\omega _{0}\left( k\right) \right]
\approx 1$. Substituting this into Eq.(\ref{Axt2}), and using the normalized
spatial frequency $q=Xk$ we get 
\begin{equation}
A^{\pm }\left( \xi ,0\right) =\frac{1}{4\pi }\sum_{m=-\infty }^{\infty
}\left( a_{m,0}\pm a_{m,1}\right) \int_{-\infty }^{+\infty }dq\,\mathcal{G}%
\left( q\right) e^{iq\left( \xi -m\right) }.
\end{equation}%
Now we need to choose a particular cut--off function. To obtain an explicit
solution of Eq. (\ref{Diff.eq.}) we adopt a Gaussian, 
\begin{equation}
\mathcal{G}\left( q\right) =e^{-w^{2}q^{2}},  \label{Gauss}
\end{equation}%
where $w$ is a free parameter. We further consider the usual case that the
QW starts with the particle at position $m=0$, and we get 
\begin{eqnarray}
A^{\pm }\left( \xi ,0\right) &=&a_{0,0}G\left( 0\right) \pm a_{-1,1}G\left(
-1\right) \pm a_{1,1}G\left( 1\right) ,  \label{initial} \\
G\left( m\right) &=&N\exp \left[ -\frac{\left( \xi -m\right) ^{2}}{4w^{2}}%
\right] ,  \notag
\end{eqnarray}%
with $N$ a normalization factor that can be fixed by impossing 
\begin{equation}
\sum_{j=+,-}\left[ \int_{-\infty }^{+\infty }d\xi \,\left\vert R^{j}\left(
\xi ,0\right) \right\vert ^{2}+\int_{-\infty }^{+\infty }d\xi \,\left\vert
L^{j}\left( \xi ,0\right) \right\vert ^{2}\right] =1,
\end{equation}%
for the total probability remains equal to unity at all times. In the
following we omit this constant, as it is nothing but a scaling factor.

In substituting the initial condition (\ref{initial}) into Eq.(\ref{Aq0})
and this into Eq.(\ref{aprel}), we are left with integrals of the form 
\begin{eqnarray}
\mathcal{Z}\left( \xi ,\tau \right) &=&\int_{-\infty }^{+\infty }dk\,\exp
\left( i\mathcal{A}k+i\frac{1}{3}\mathcal{B}k^{3}-\mathcal{C}k^{2}\right) ,
\label{zeta} \\
\mathcal{A} &=&\xi -\sqrt{\rho }\tau , \\
\mathcal{B} &=&\frac{1}{2}\sqrt{\rho }\left( 1-\rho \right) \tau , \\
\mathcal{C} &=&w^{2},
\end{eqnarray}%
whose solution can be expressed in terms of the Airy function $A_{i}\left(
x\right) $ \cite{Miyagi,Abramowitz} 
\begin{equation}
\mathcal{Z}\left( \xi ,\tau \right) =\frac{2\pi }{\mathcal{B}^{1/3}}\exp
\left( \frac{3\mathcal{ABC}+2\mathcal{C}^{3}}{3\mathcal{B}^{2}}\right)
A_{i}\left( \frac{\mathcal{AB}+\mathcal{C}^{2}}{\mathcal{B}^{4/3}}\right) .
\end{equation}

Then the final result reads, up to a normalization factor, 
\begin{eqnarray}
\bar{a}\left( \xi ,\tau \right)  &=&A^{+}\left( \xi ,\tau \right) +\left(
-1\right) ^{n}A^{-}\left( \xi ,\tau \right)   \label{sol} \\
A^{\pm }\left( \xi ,\tau \right)  &=&a_{0,0}\mathcal{Z}\left( \pm \xi ,\tau
\right) \pm a_{-1,1}\mathcal{Z}\left( \pm \left( \xi +1\right) ,\tau \right)
\pm a_{1,1}\mathcal{Z}\left( \pm \left( \xi -1\right) ,\tau \right) ,  \notag
\end{eqnarray}%
or, explicitely, 
\begin{eqnarray}
\bar{R}\left( \xi ,\tau \right)  &=&R^{+}\left( \xi ,\tau \right) +\left(
-1\right) ^{n}R^{-}\left( \xi ,\tau \right)   \label{solfin1} \\
\bar{L}\left( \xi ,\tau \right)  &=&L^{+}\left( \xi ,\tau \right) +\left(
-1\right) ^{n}L^{-}\left( \xi ,\tau \right)   \label{solfin2}
\end{eqnarray}%
with 
\begin{eqnarray}
R^{\pm }\left( \xi ,\tau \right)  &=&R_{0,0}\left[ \mathcal{Z}\left( \pm \xi
,\tau \right) \pm \sqrt{\rho }\mathcal{Z}\left( \pm \left( \xi -1\right)
,\tau \right) \right]  \\
&&\pm \sqrt{1-\rho }L_{0,0}\mathcal{Z}\left( \pm \left( \xi +1\right) ,\tau
\right) ,  \notag \\
L^{\pm }\left( \xi ,\tau \right)  &=&L_{0,0}\left[ \mathcal{Z}\left( \pm \xi
,\tau \right) \mp \sqrt{\rho }\mathcal{Z}\left( \pm \left( \xi +1\right)
,\tau \right) \right]  \\
&&\pm \sqrt{1-\rho }R_{0,0}\mathcal{Z}\left( \pm \left( \xi -1\right) ,\tau
\right) ,  \notag
\end{eqnarray}%
where use was made of the discrete QW Eqs.(\ref{amp1}) and (\ref{amp2}). 

Notice that while the form of Eq. (\ref{Diff.eq.}) is independent of the
particular form of $\mathcal{G}\left( kX\right) $, this solution depends
explictly on $w$ because the application of the initial condition does
depend on the details of that cut-off function.

Finally, the probability distribution is given by 
\begin{eqnarray}
P\left( \xi ,\tau \right) &=&P^{R}\left( \xi ,\tau \right) +P^{L}\left( \xi
,\tau \right) , \\
P^{A}\left( \xi ,\tau \right) &=&\left\vert A^{+}\left( \xi ,\tau \right)
+\left( -1\right) ^{n}A^{-}\left( \xi ,\tau \right) \right\vert
^{2},\;\;A=R,L.  \notag
\end{eqnarray}

Let us now compare the above long wavelength approximation with the solution
of the original discrete QW Eqs.(\ref{amp1}) and (\ref{amp2}). In the left
column of Fig.2 we present the total probability distribution $P_{m}\left(
n\right) $ , as well as the partial probabilites $P_{m}^{R}\left( n\right) $
and $P_{m}^{L}\left( n\right) $, as obtained from Eqs.(\ref{amp1}) and (\ref%
{amp2}) for $n=200$, $R_{0,0}=1/\sqrt{2}$ and $L_{0,0}=i/\sqrt{2}$ (a value
of the initial conditions for which the mean displacement is null) and $\rho
=1/2$ (which corresponds to the Hadamard walk). In the right column of Fig.2
we present the solution of the long wavelength approximation $P\left( \xi
,\tau \right) $ (and $P^{R}\left( \xi ,\tau \right) \ $and $P^{L}\left( \xi
,\tau \right) $) for $\tau =200$ and the same value of $\rho $ and of the
initial conditions and for $w=0.4$. Finally, in the central column of the
figure we present the long wavelength approximation evaluated only at
integer values of the position $\xi $. These last plots are given in order
to make a better comparison with the exact solution, as the latter is
nonvanishing only at even (odd) positions at even (odd) times. Although we
only show plots for a particular initial condition and value of $\rho $,
similar correspondence between the exact solution and the long wavelength
limit results for other sets of initial conditions and values of $\rho $ as
well.

In Fig. 2 we chose the value $w=0.4$ because for that particular value the
agreement of the long wavelength approximation with the exact solution is
close to the best that can be achieved. In Fig. 3 we plot the long
wavelength approximation for three different values of $w$. Note that for
the largest $w=0.55$ the long wavelength approximation yields values that
are too low, compared to the exact solution, for small $\xi $, while for the
smallest $w=0.25$ the long wavelength approximation extends to $\left\vert
\xi \right\vert $ beyond the range of the exact solution.

We see from Fig. 2 that the main trends of the exact QW solution are well
captured by the long wavelength approximation, although obvious quantitative
differences appear. The main discrepancy between the exact solution and the
long wavelength approximation is the presence of the oscillations that
appear in the latter. By taking into account higher order terms in the
expansion (\ref{series}) a more accurate approximation to the exact solution
could be obtained, but the corresponding differential equation would not
have a simple analytic solution. In any case, our main thesis is not that
Eqs. (\ref{solfin1}) and (\ref{solfin2}) are a good quantitative
approximation to the exact solution. Rather, our central result is that
comparisons such as those in Fig. 2 demonstrate that the propagation and
dispersion effects described by Eq. (\ref{Diff.eq.}) capture the qualitative
nature of the QW.

\section{Conclusions}

In this paper we have rigorously derived both the exact solution of the
coined QW for arbitrary unitary transformations, and a long wavelength
approximation that constitutes an approximate continuous limit of the QW.

The long wavelength approximation identified here corresponds to the
dynamical equation of an envelope function characterizing a light pulse
propagating in a medium with higher order dispersion, for example in an
optical fiber close to the zero-dispersion wavelength \cite{Agrawal}. This,
together with the fact that entirely classical implementations of the
unidimensional QW are possible \cite{Knight,Knight2}, lead us to the
conclusion that entanglement does not play a central role in the
unidimensional QW, and that the walk can be regarded as a pure wave
phenomenon. This conclusion cannot, however, be generalized to higher
dimensional QWs \cite{Mackay}. There several coins are needed, and it would
in such a scenario that the entanglement characteristic of an essential
quantum nature might appear, as already discussed \cite{Knight2}. It would
then be very interesting to generalize the work we have carried out here by
deriving long wavelength descriptions of multidimensional QWs. This could
help elucidate any role played by entanglement in such processes.

Another point of interest is the effect of decoherence on QWs. Kendon and
Tregenna \cite{Kendon} have shown that decoherence modifies the probability
distribution of the QW in somewhat unsuspected ways. In particular, it can
lead to quite smooth, and at the same time wide, probability distributions.
It would be interesting to investigate the role of decoherence in continuous
descriptions of the QW, such as that we have presented here. In this sense,
the recent work by Romanelli \textit{et al.} \cite{Romanelli}, in which
Markovian and intereference terms of the quantum evolution are separated,
provides a new interesting tool for this type of analysis.

\section{Acknowledgements}

This work has been supported in part by the UK Engineering and Physical
Sciences Research Council and the European Union. ER acknowledges financial
support from the Ministerio de Educaci\'{o}n, Cultura y Deportes of the
Spanish Goverment (Grant PR20002-0244). JES acknowledges financial support
from the Natural Sciences and Engineering Research Council of Canada. We
gratefully acknowledge fruitful discussions with V. Kendon.

\section{Appendix}

In this appendix we show the relation between the walk definition we use
here and that used by Nayak and Vishwanath\textit{\ }\cite{Nayak}, and
confirm that our solution reproduces theirs. We begin with our walk in the
case of a Hadamard transformation, 
\begin{eqnarray}
R_{m,n} &=&\frac{1}{\sqrt{2}}\left( R_{m-1,n-1}+L_{m+1,n-1}\right) ,
\label{Hadamard1} \\
L_{m,n} &=&\frac{1}{\sqrt{2}}\left( R_{m-1,n-1}-L_{m+1,n-1}\right) .
\label{Hadamard2}
\end{eqnarray}%
and define new variables 
\begin{equation}
\hat{R}_{m,n}\equiv L_{-m+1,n},\ \ \hat{L}_{m,n}\equiv R_{-m-1,n},
\label{Nfromus}
\end{equation}%
so 
\begin{equation}
L_{m,n}=\hat{R}_{-m+1,n},\ \ R_{m,n}=\hat{L}_{-m-1,n.}  \label{usfromN}
\end{equation}%
In terms of these new variables the walk Eqs.(\ref{Hadamard1}) and (\ref%
{Hadamard2}) are specified by 
\begin{eqnarray}
\hat{L}_{m,n} &=&\frac{1}{\sqrt{2}}\left( \hat{L}_{m+1,n-1}+\hat{R}%
_{m+1,n-1}\right) , \\
\hat{R}_{m,n} &=&\frac{1}{\sqrt{2}}\left( \hat{L}_{m-1,n-1}+\hat{R}%
_{m-1,n-1}\right) .
\end{eqnarray}%
These are the walk equations of Nayak and Vishwanath \cite{Nayak}\textit{. }%
Then, in terms of our functions $a^{\pm }(x,t)$, where $a$ is $R$ or $L$, we
have from Eqs.(\ref{finres}) and (\ref{Nfromus}), 
\begin{eqnarray}
\hat{R}_{m,n} &=&L^{+}((-m+1)X,nT)+(-1)^{n}L^{-}((-m+1)X,nT), \\
\hat{L}_{m,n} &=&R^{+}((-m-1)X,nT)+(-1)^{n}R^{-}((-m-1)X,nT).
\end{eqnarray}%
In writing out the expressions for $R^{\pm }$ and $L^{\pm }$ at the lattice
sites indicated we do a change of variables, putting $k=k^{\prime }+\pi /X$.
We use 
\begin{eqnarray}
\omega _{0}(k^{\prime }+\pi /X) &=&-\omega _{0}(k^{\prime }), \\
\exp \left[ i(k^{\prime }+\pi /X)X\right]  &=&-\exp \left[ ik^{\prime }X%
\right] , \\
\exp \left[ -i(k^{\prime }+\pi /X)mX\right]  &=&(-1)^{m}\exp \left[
-ik^{\prime }mX\right] ,
\end{eqnarray}%
and can then adjust the limits of integration of the function back to $-\pi
/X$ to $\pi /X$ in the new variable $k^{\prime }$, since the function is
periodic over $2\pi /X$. Changing the name of the new variable back to $k$
we can collect the terms and write 
\begin{eqnarray}
\hat{R}_{m,n} &=&\int_{-\pi /X}^{\pi /X}dk\,e^{ikX}\left[
L^{+}(k)-(-1)^{n+m}L^{-}(k+\frac{\pi }{X})\right]   \notag \\
&&\cdot e^{-ikmX}e^{-i\omega _{0}(k)nT}, \\
\hat{L}_{m,n} &=&\int_{-\pi /X}^{\pi /X}dk\,e^{-ikX}\left[
R^{+}(k)-(-1)^{n+m}R^{-}(k+\frac{\pi }{X})\right]   \notag \\
&&\cdot e^{-ikmX}e^{-i\omega _{0}(k)nT}.
\end{eqnarray}%
Now the initial condition adopted in\textit{\ }\cite{Nayak} is $\hat{L}%
_{m,0}=\delta _{m0}$ and $\hat{R}_{m,0}=0$, which, in our notation Eqs.(\ref%
{usfromN}) read $R_{m,0}=\delta _{m,-1}$ and $L_{m,0}=0.$ After one time
step, see Eqs.(\ref{Hadamard1}) and (\ref{Hadamard2}), we have $%
R_{m,1}=L_{m,1}=\delta _{m0}/\sqrt{2}$, so we can construct $L^{\pm }(k)$
and $R^{\pm }(k)$ from (\ref{a(k)}), with $\rho =1/2$. We find 
\begin{eqnarray}
L^{+}(k) &=&-L^{-}(k+\frac{\pi }{X})=\frac{X}{4\pi }\frac{1}{\sqrt{1+\cos
^{2}kX}}, \\
R^{+}(k) &=&-R^{-}(k+\frac{\pi }{X})=\frac{X}{4\pi }\frac{\sqrt{2}e^{i\omega
_{0}(k)T}e^{ikX}+1}{\sqrt{1+\cos ^{2}kX}},
\end{eqnarray}%
and using (\ref{disp}) we recover 
\begin{eqnarray}
\hat{R}_{m,n} &=&\left[ 1+(-1)^{n+m}\right] \frac{X}{4\pi }  \notag \\
&&\cdot \int_{-\pi /X}^{\pi /X}dk\frac{e^{ikX}}{\sqrt{1+\cos ^{2}kX}}%
e^{-ikmX}e^{-i\omega _{0}(k)nT}, \\
\hat{L}_{m,n} &=&\left[ 1+(-1)^{n+m}\right] \frac{X}{4\pi }  \notag \\
&&\cdot \int_{-\pi /X}^{\pi /X}dk\left( 1+\frac{\cos kX}{\sqrt{1+\cos ^{2}kX}%
}\right) e^{-ikmX}e^{-i\omega _{0}(k)nT},
\end{eqnarray}%
in agreement with Nayak and Vishwanath \cite{Nayak}

\bigskip

{\LARGE Figure Captions\bigskip }

\textbf{Fig.1.} Dispersion relations as given by Eq.(\ref{disp})

\textbf{Fig.2.} Probability distribution of the QW. The exact solution is
presented in the left column ($P_{m}\left( n\right) $, $P_{m}^{R}\left(
n\right) $ and $P_{m}^{L}\left( n\right) $ for $n=200$ from top to bottom),
the long wavelength approximation ($P\left( \xi ,\tau \right) $, $%
P^{R}\left( \xi ,\tau \right) $, and $P^{L}\left( \xi ,\tau \right) $ for $%
\tau =200$ from top to bottom) is presented in the right column, and also in
the central column but in this case evaluated only at integer even. In the
left and central columns the points have been joined to guide the eye. The
initial condition is $R_{0,0}=1/\sqrt{2}$ and $L_{0,0}=i/\sqrt{2}$, and the
parameters are $\rho =1/2$, (Hadamard walk) and $w=0.4$ for the long
wavelength approximation. The probabilities in the long wavelength
approximation have not been normalized, and the units are thus arbitrary.

\textbf{Fig.3.} $P\left( \xi ,\tau \right) $ for $\tau =200$ and the same
initial conditions and parameters as in Fig.3 except for the values of $w$
that are the indicated in the figure.

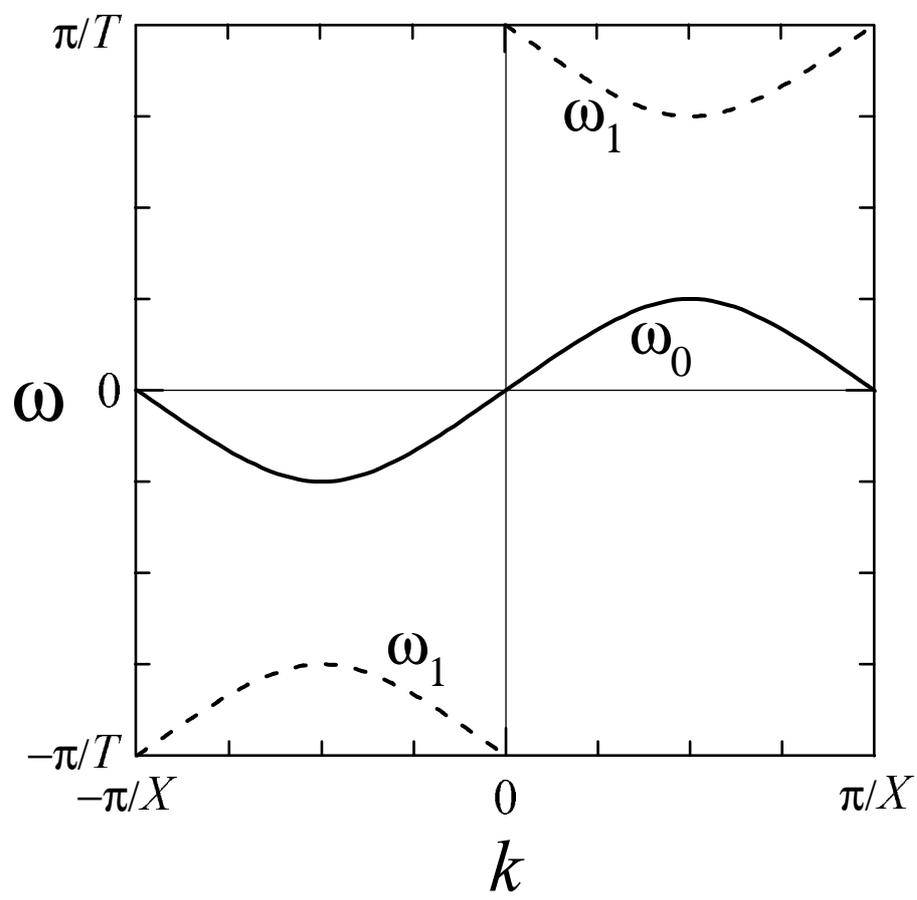
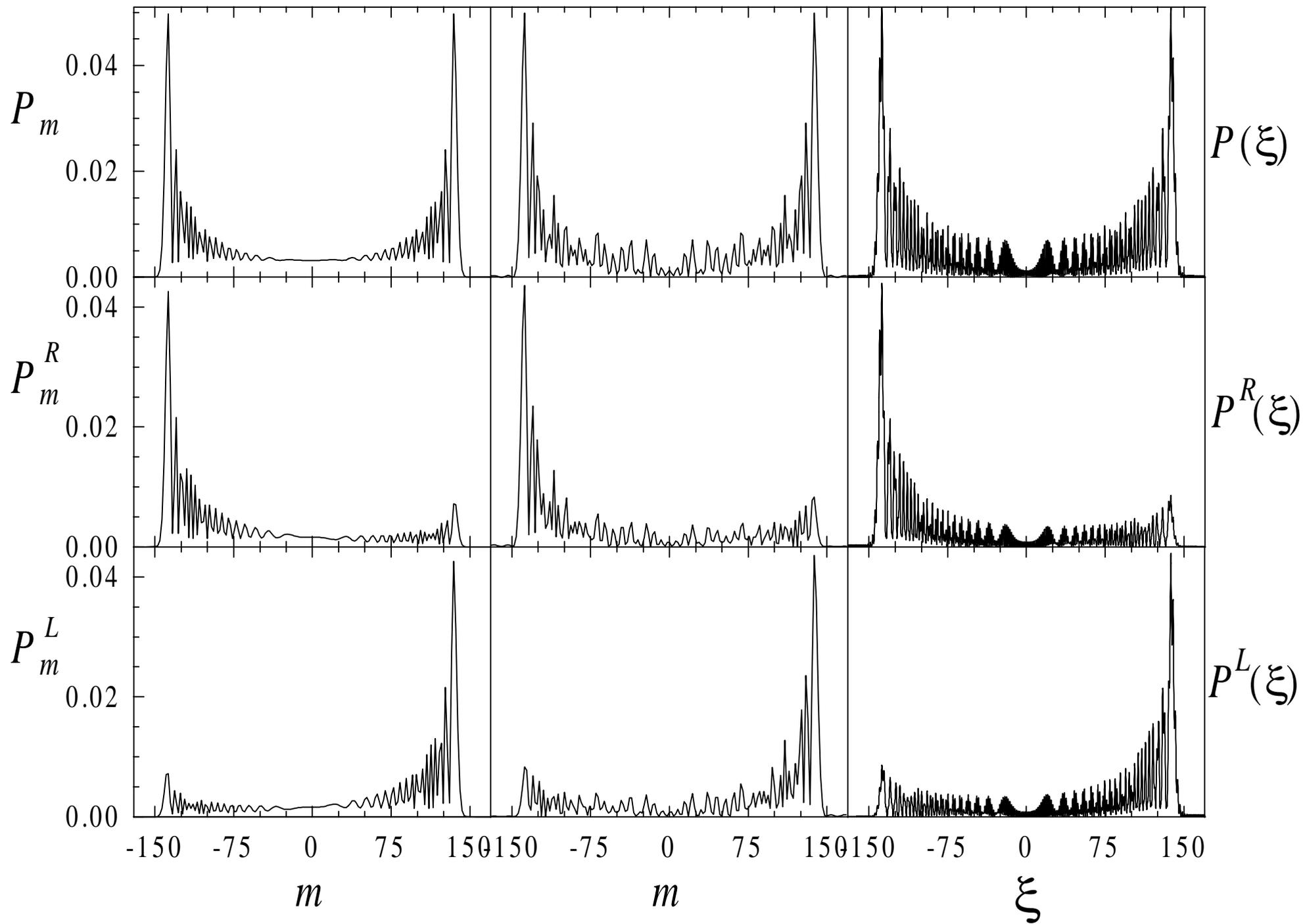
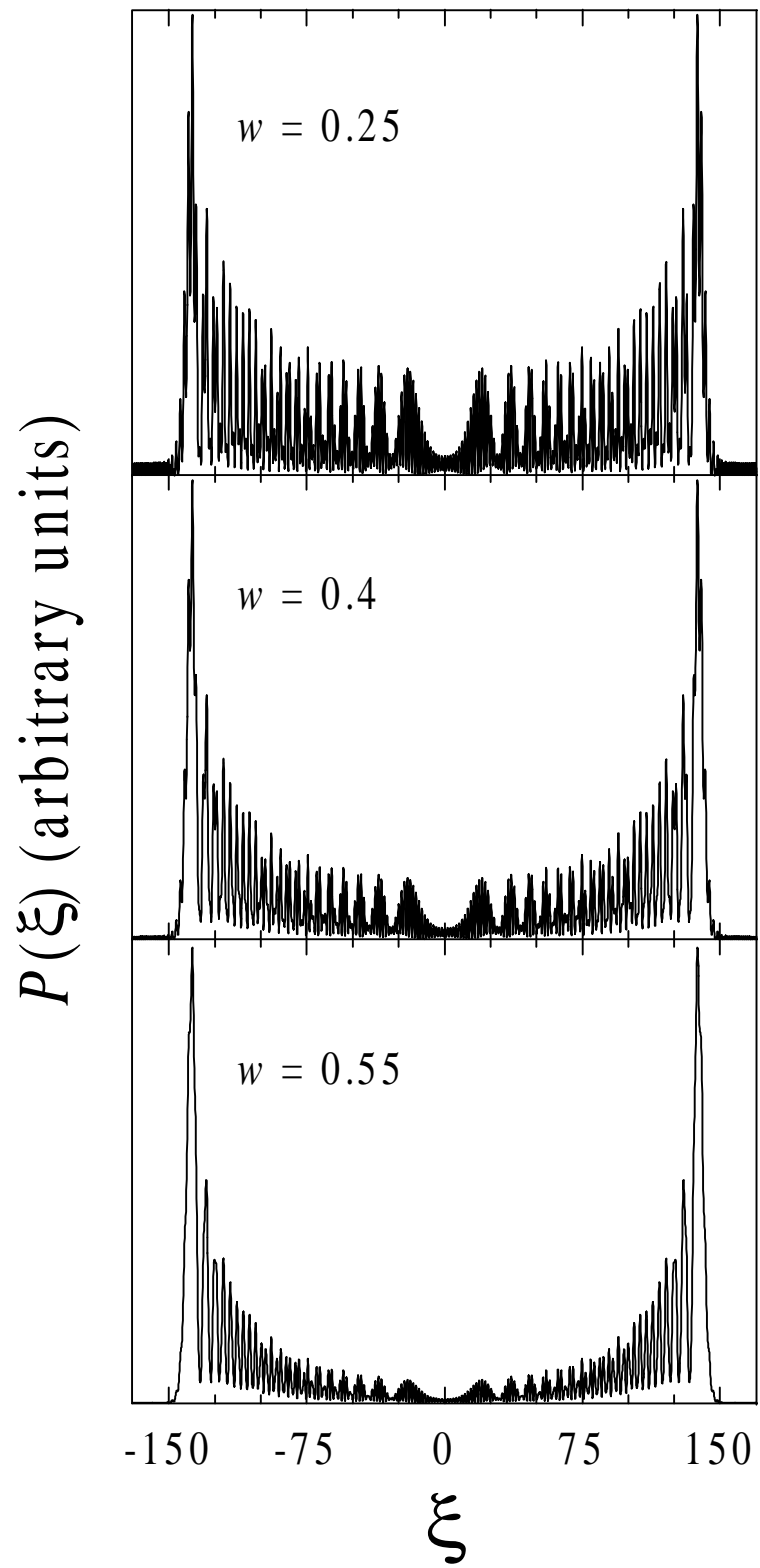
\end{document}